\documentclass[aip,jcp,reprint,amsmath,amssymb,floatfix,citeautoscript,nofootinbib]{revtex4-2}
\usepackage{cancel}
\usepackage{amsmath}
\usepackage{amssymb} 
\usepackage{amsthm}
\usepackage{bm}
\usepackage{physics}
\usepackage{graphicx}
\usepackage{multirow}
\usepackage{ragged2e}
\usepackage{newtxtext,newtxmath}
\usepackage{comment}
\allowdisplaybreaks

\usepackage{xr}
\usepackage{color}
\usepackage[usenames,dvipsnames]{xcolor}
\definecolor{myblue}{rgb}{0,0,1}
\usepackage[breaklinks=true,colorlinks=true,linkcolor=myblue,urlcolor=myblue,citecolor=myblue]{hyperref}
\usepackage{cleveref}

\usepackage{array}[=2016-10-06]
\usepackage{placeins}
\usepackage{booktabs}
\usepackage{tabularx}
\usepackage[version=4]{mhchem}
\newcolumntype{R}[1]{>{\raggedright\arraybackslash}p{#1}}
\newcolumntype{L}[1]{>{\raggedleft\arraybackslash}p{#1}}
\newcolumntype{C}[1]{>{\centering\arraybackslash}p{#1}}
\newcolumntype{M}[1]{>{\centering\arraybackslash}m{#1}}

\newcommand{\mbf}[1]{\mathbf{#1}}
\newcommand{\mrm}[1]{\mathrm{#1}}

\begin{document}

\title{Active-space construction in single-fragment coupled-cluster embedding for chemical reactions}

\author{Ardavan Farahvash}
\affiliation{Department of Chemistry, Columbia University, New York, New York 10027, USA}
\author{Timothy C. Berkelbach}
\email{t.berkelbach@columbia.edu}
\affiliation{Department of Chemistry, Columbia University, New York, New York 10027, USA}
\affiliation{Initiative for Computational Catalysis, Flatiron Institute, New York, New York 10010, USA}

\begin{abstract}
We benchmark single-fragment quantum embedding, in which electron correlation---at the level of coupled-cluster theory with single, double, and perturbative triple excitations [CCSD(T)]---is restricted to orbitals associated with a chosen molecular fragment, and the remaining orbitals are frozen at the Hartree–Fock level. 
We compare several methods for constructing fragment active spaces, and we test their performance on 60 organic reaction energies and three reaction profiles. 
A common fragment projection operator is used to relate several of the tested orbital constructions.
Regional embedding, based on density matrix embedding theory, is shown to give accurate and robust performance when used for both the occupied and virtual active orbitals.
For virtual orbitals, fragment-centered projected atomic orbitals and fragment natural orbitals (defined by eigenvectors of a fragment MP2 density matrix) also perform well, but the latter typically retains larger virtual spaces.
Composite MP2 corrections yield the largest improvement: combined with fragment natural orbitals, they give average reaction energy errors below 0.3 kcal/mol relative to full-system CCSD(T), with all individual errors below 1 kcal/mol.
Reaction profiles show that accurate endpoint energies can conceal substantial barrier errors, emphasizing the importance of approximating omitted correlation.
\end{abstract}
 
\maketitle
\section{Introduction}

The applicability of high-level quantum chemistry methods is often hampered by their steep computational scaling. 
Quantum embedding methods seek to alleviate this bottleneck by partitioning a system into two regions, the ``fragment'' and the ``environment'', where the fragment (e.g., a reaction center) is treated using a high level of quantum theory, and the environment is treated using a low level of quantum theory~\cite{Libisch2014,Sun2016,Lee2019,Jones2020}. 
Here, we consider a particularly simple version of this idea: a single-fragment, one-shot embedding in which a global Hartree–Fock (HF) calculation is followed by a correlated calculation in a restricted set of occupied and virtual orbitals associated with a chemically important fragment.

Our focus is on coupled-cluster theory with single, double, and perturbative triple excitations [CCSD(T)], which is a successful method for main-group chemistry~\cite{Bartlett2007}.
Recovering dynamic correlation via CCSD(T) requires large active spaces, motivating modifications to standard active-space construction schemes that typically target static correlation in a small active space~\cite{Toth2020}.
Although any single-fragment calculation provides a poor approximation to the total energy, good performance is achievable for energy differences, as long as the errors cancel.
To do so, the active spaces must be balanced across geometries.

In this work, we test how different active-space constructions affect accuracy of chemical reaction energies and barrier heights.
We compare methods based on various types of local orbitals and their assignment to fragments, using a projection operator to unify most of them.
Finally, we assess the efficacy of composite corrections based on second-order perturbation theory to include electron correlation beyond the fragment.

\section{Theory}
\label{sec:theory}

\subsection{Coupled-cluster in Hartree-Fock embedding}

The active-space quantum embedding methods considered in this work exploit the fact that a Slater determinant is unchanged by separate unitary mixings of the occupied and virtual molecular orbitals (MOs).
Following a global HF calculation, we rotate the occupied and virtual orbitals,
\begin{subequations}
\begin{align}
    |\tilde{i}\rangle &= \sum_i U_{i\tilde{i}} |i\rangle, \\
    |\tilde{a}\rangle &= \sum_a V_{a\tilde{a}} |a\rangle.
\end{align}
\end{subequations}
A subset of these orbitals is assigned to the fragment and retained as the active space, and the rest are frozen.
We semi-canonicalize the active-space orbitals and perform a CCSD(T) calculation, yielding the total energy
\begin{equation}
    \label{eq:cc-in-hf}
    E_{\text{CC-in-HF}}
    =
    E_0^{(\mathrm{HF})}
    +
    E_{\text{c,act}}^{(\mathrm{CC})},
\end{equation}
where $E_0^{(\mathrm{HF})}$ is the HF energy of the full system and $E_{\text{c,act}}^{(\mathrm{CC})}$ is the correlation energy obtained by correlating electrons only within the active space.
Equivalently, the frozen occupied orbitals contribute to the active-space Hamiltonian through a constant energy and an effective one-electron operator,
\begin{equation}
    H_\mathrm{act}
    =
    E_\mathrm{frz}
    +
    \sum_{pq}^{\mrm{act}}
    h_{pq}^{(\mathrm{eff})}a_p^\dagger a_q
    +
    \frac{1}{2}
    \sum_{pqrs}^{\mrm{act}}
    \langle pq|rs\rangle
    a_p^\dagger a_q^\dagger a_s a_r ,
\end{equation}
with
\begin{align}
    E_\mrm{frz}
    &=
    \sum_i^\mrm{frz} h_{ii}
    +
    \sum_{ij}^\mrm{frz}
    \left(
    \langle ij|ij\rangle-\langle ij|ji\rangle
    \right), \\
    h_{pq}^{(\mathrm{eff})}
    &=
    h_{pq}
    +
    \sum_i^\mrm{frz}
    \left[
    \langle pi|qi\rangle-\langle pi|iq\rangle
    \right].
\end{align}
We will consider several methods to generate active occupied orbitals, virtual orbitals, or both, as described in the following sections.

\subsection{Fragment projection operator}

Several of the orbital constructions considered here can be expressed in terms of a projection operator onto an atomic or fragment subspace.
For an atom $A$, we define the atomic projector
\begin{equation}
    \hat{P}_A
    =
    \sum_{\mu' \in A}
    |\mu'\rangle\langle\mu'|,
\end{equation}
where $\{|\mu'\rangle\}$ is an orthonormal set of atomic orbitals (AOs), of which there are multiple possible choices.
For a fragment $F$ containing several atoms, we define the fragment projector
\begin{equation}
    \hat{P}_F
    =
    \sum_{A\in F}\hat{P}_A.
\end{equation}
The methods below differ primarily in the choice of orthonormal AOs, in whether the rotated orbitals are obtained iteratively or noniteratively, and in how the resulting orbitals are selected for the active space.

\subsection{Localized molecular orbitals}

We first consider conventional orbital-localization procedures~\cite{Lehtola2013}, including Boys orbitals, which minimize
\begin{equation}
    L_{\text{Boys}}
    =
    \sum_{\tilde{i}}
    \mel{\tilde{i}}{\mbf{r}}{\tilde{i}}^2,
\end{equation}
and Pipek--Mezey (PM) orbitals, which maximize 
\begin{equation}
    L_{\text{PM}}
    =
    \sum_A
    \sum_{\tilde{i}}
    \left[
    \langle\tilde{i}|\hat{P}_A|\tilde{i}\rangle
    \right]^2 .
\end{equation}
For PM localization, we test two choices for the atomic orbitals $|\mu'\rangle$:
meta-L\"owdin orbitals~\cite{sun_exact_2014} and intrinsic atomic orbitals (IAOs)~\cite{knizia_intrinsic_2013}, the latter of which generate so-called intrinsic bond orbitals (IBOs) when used in PM localization.

Once the orbitals have been localized by any of the above procedures, an orbital is assigned to the fragment active space when its total fragment population exceeds a threshold,
\begin{equation}
    \langle\tilde{i}|\hat{P}_F|\tilde{i}\rangle > \epsilon.
\end{equation}
Although most effective for occupied orbitals, we also test these methods for localization and truncation of the virtual orbitals.

\subsection{Projected atomic orbitals}

For the virtual space, we also consider projected atomic orbitals (PAOs)~\cite{Pulay1983},
\begin{equation}
    |r_\mu\rangle
    =
    \left(
    1-\sum_i |i\rangle\langle i|
    \right)
    |\mu\rangle .
\end{equation}
We consider two ways of associating PAOs with the fragment.
Following local correlation methods~\cite{boughton_comparison_1993,hampel_local_1996}, PAOs are retained according to their fragment population,
\begin{equation}
    \langle r_\mu|\hat{P}_F|r_\mu\rangle>\epsilon.
\end{equation}
Alternatively, we retain only PAOs generated from AOs belonging to the fragment, $\mu\in F$; we refer to these as FragPAOs, which conveniently have no associated thresholds.
Because PAOs are neither orthogonal nor linearly independent, the selected PAO space is canonically orthogonalized with linear dependencies removed (overlap eigenvalues below $10^{-6}$).

\subsection{Projection-based fragment orbitals: DMET and regional embedding}

An alternative route to fragment orbitals is obtained by diagonalizing the fragment projection operator directly within the occupied and virtual MO spaces,
\begin{subequations}
\begin{align}
    \mathbf{P}_F^\mrm{O}\mathbf{U}
    &=
    \mathbf{U}\,\mathrm{diag}(n_{\tilde{i}}), \\
    \mathbf{P}_F^\mrm{V}\mathbf{V}
    &=
    \mathbf{V}\,\mathrm{diag}(n_{\tilde{a}}),
\end{align}
\end{subequations}
where
\begin{subequations}
\begin{align}
    [\mathbf{P}_F^\mrm{O}]_{ij}
    &=
    \langle i|\hat{P}_F|j\rangle, \\
    [\mathbf{P}_F^\mrm{V}]_{ab}
    &=
    \langle a|\hat{P}_F|b\rangle.
\end{align}
\end{subequations}
The eigenvectors define rotated occupied and virtual orbitals, and the eigenvalues $n_{\tilde{i}}$ and $n_{\tilde{a}}$ give their fragment populations.
For a fragment projector of rank $N_F$, there are at most $N_F$ occupied and $N_F$ virtual orbitals with nonzero fragment population, so the corresponding occupied-virtual active space contains at most $2N_F$ orbitals.
In practice, we correlate orbitals whose fragment population exceeds a threshold, which is different for the occupied and virtual spaces.
Unlike conventional localization, which attempts to associate orbitals with individual atoms or bonds, this construction directly identifies linear combinations with maximal support on the entire fragment.

This occupied-virtual formulation is closely related to the more familiar fragment-bath formulation of density matrix embedding theory (DMET)~\cite{Knizia2012,knizia_density_2013}.
To see the connection, consider an eigenvector of the occupied-space fragment projector with eigenvalue $0<n_i<1$.
The corresponding occupied orbital can be written as
\begin{equation}
    |\tilde{i}\rangle
    =
    \sqrt{n_i}\,|f_i\rangle
    +
    \sqrt{1-n_i}\,|b_i\rangle,
\end{equation}
where $|f_i\rangle$ and $|b_i\rangle$ are normalized components within the fragment and ``bath'' (i.e., non-fragment) orbital subspaces, respectively.
Within the same two-dimensional subspace, the orthogonal complementary orbital is
\begin{equation}
    |\tilde{a}_i\rangle
    =
    \sqrt{1-n_i}\,|f_i\rangle
    -
    \sqrt{n_i}\,|b_i\rangle.
\end{equation}
Because $|\tilde{a}_i\rangle$ is orthogonal to the occupied space, it lies in the virtual space, and its fragment population is
\begin{equation}
    n_{\tilde{a}_i}=1-n_i.
\end{equation}
Therefore,
\begin{equation}
    \mathrm{span}
    \left\{
    |f_i\rangle,|b_i\rangle
    \right\}
    =
    \mathrm{span}
    \left\{
    |\tilde{i}\rangle,|\tilde{a}_i\rangle
    \right\}.
\end{equation}
Thus, the conventional DMET fragment--bath orbitals and the corresponding occupied--virtual orbitals are merely different bases for the same one-particle embedding space, as first discussed in the Appendix of Ref.~\citenum{zheng_ground-state_2016}.

    \begin{figure}[tb]
        \centering
        \includegraphics[width=\columnwidth]{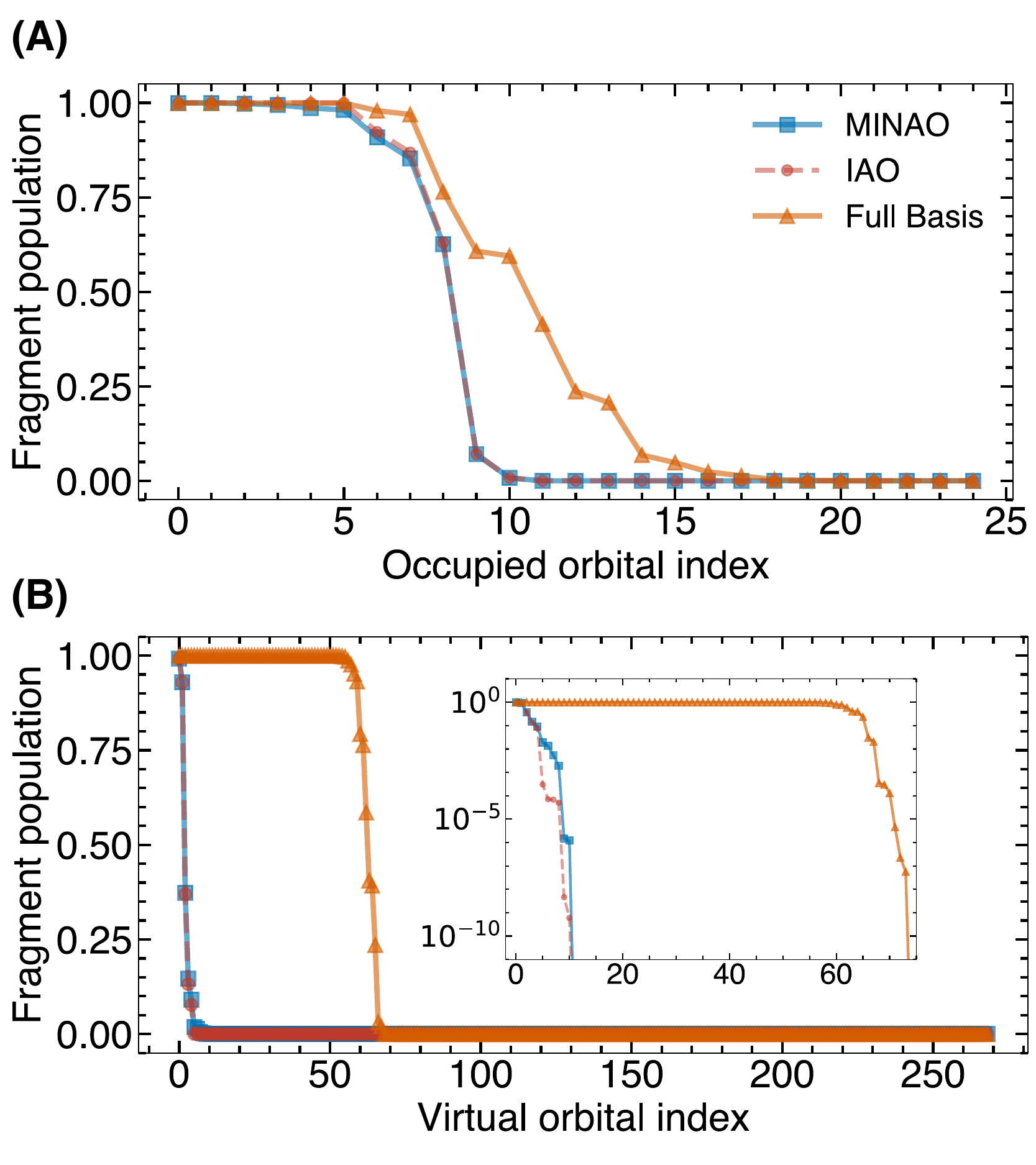}
        \caption{Fragment populations of the phenol molecule in the cc-pVTZ basis set using regional embedding (A) occupied orbitals and (B) virtual orbitals, using various fragment AO basis sets. The inset in (B) plots the same curve on a log-scale.}
        \label{fig:intro2}
    \end{figure}

In the calculations presented here, we use the regional embedding construction of Ref.~\citenum{lau_regional_2021}, which generalizes the above idea by allowing different fragment projectors for the occupied and virtual spaces; throughout the work, we refer to this method as ``DMET-RE''.
For the occupied space, we use a valence-only minimal atomic orbital basis on the fragment (specifically, the so-called MINAO basis that retains only the valence shells of the cc-pVTZ basis); for the virtual space, we use the full computational basis on the fragment.
The former provides a compact description of the fragment valence occupied space, whereas the latter retains the much larger virtual space needed to describe dynamical correlation.
We also tested the use of IAOs for the occupied space projector, and found the results were nearly identical to those obtained using MINAOs.
In Fig.~\ref{fig:intro2}, we show the fragment populations of the DMET-RE orbitals (occupied and virtual), comparing the use of MINAO and full basis (cc-pVTZ) projectors: clearly, the difference for the occupied orbitals is relatively minor, whereas the difference for the virtual orbitals is significant.

    \begin{figure*}[t]
        \centering
        \includegraphics[width=0.6\textwidth]{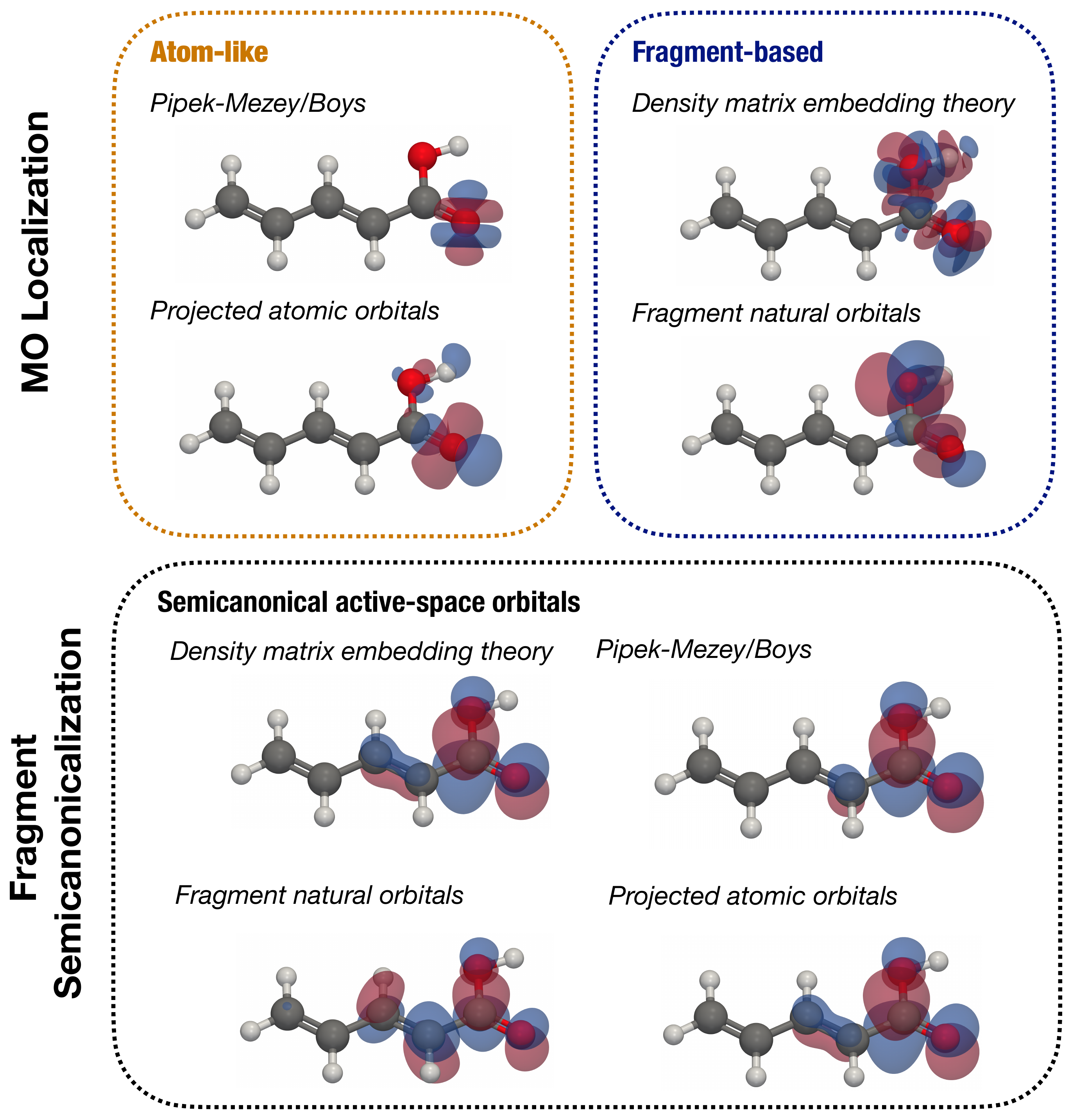}
        \caption{Examples of localized MOs and semicanonical active-space orbitals for penta-2,4-dienoic acid. 
        The fragment atoms are chosen to be those of the carboxyl group. The orbitals shown in the top panels are all localized virtual orbitals found via a rotation of the canonical virtual orbitals. The orbitals shown in the bottom panel are semicanonical LUMO+1 orbitals.
        }
        \label{fig:intro1}
    \end{figure*}
    
The projector construction presented above also clarifies the close relationship between several active-space embedding methods.
For example, the atomic valence active space~\cite{sayfutyarova_automated_2017}, the subsystem-projected AO decomposition (SPADE)~\cite{claudino_automatic_2019,claudino_simple_2019}, and active-space embedding theory~\cite{He2020} differ primarily in the subspace used to define $\hat{P}_F$ and in the criterion used to select orbitals for the active space.
For example, the SPADE approach~\cite{claudino_automatic_2019,claudino_simple_2019} recommends truncating the fragment space at the largest inflection point in the ordered fragment populations.
We will compare this truncation criterion against the use of fixed thresholds in some of our results below.

\subsection{Fragment natural orbitals}
\label{subsec:fragnos}

The final class of active orbitals considered here are fragment natural orbitals (FragNOs), inspired by local natural-orbital CC methods~\cite{Rolik2011,Rolik2013,Nagy2024}.
Starting from a chosen set of fragment occupied orbitals, we evaluate their contribution to the MP2 one-particle density matrix.
Diagonalization of the corresponding occupied and virtual blocks defines the FragNOs, and the associated eigenvalues provide an importance measure for truncating the orbital space.

For the occupied space, the initially selected fragment occupied orbitals are retained explicitly, and the remaining occupied FragNOs are selected according to their MP2 occupation numbers.
The virtual active space is constructed directly from the virtual FragNOs using an analogous occupation threshold.
In this way, the FragNO construction uses approximate correlation information to identify additional orbitals that are important for describing correlation associated with the fragment.

\subsection{MP2 corrections}

For small fragments, the accuracy of the CCSD(T)-in-HF energy differences may be limited by the neglect of environmental correlation.
We can approximate this environmental correlation at an intermediate level of theory, such as second-order M\o ller-Plesset perturbation theory (MP2), which amounts to the addition of a composite correction to the energy,
\begin{equation}
    \label{eq:delmp2}
    \Delta E^{(2)} =
    E_{\mathrm{c}}^{(2)}
    -
    E_{\mathrm{c,act}}^{(2)}.
\end{equation}

\subsection{Comparing final fragment orbital spaces}
\label{ssec:comparing_fragments}

As described above, once all localized occupied and virtual orbitals are assigned to the fragment, they are semi-canonicalized.
A comparison of several different orbitals obtained by the various localization schemes is shown in Fig.~\ref{fig:intro1}, along with one of the semicanonical orbitals.
As can be seen, the semicanonical orbitals are quite similar, although there are slight differences at the fragment-environment boundary.
Nonetheless, these results suggest that the fragment active orbitals span similar valence spaces even though they were selected by different metrics.

\section{Methods}
\label{sec:methods}

\begin{table*}[t!]
    \centering
\begin{tabular*}{\textwidth}{@{\extracolsep{\fill}}lll@{}}
        \toprule
        Reaction name & Reaction formula & Number of carbons \\
        \midrule
        Alkane substitution &
        $\ce{C_n}\{\ce{CH_2Cl}\}+\{\ce{OH^-}\}\to
        \ce{C_n}\{\ce{CH_2OH}\}+\{\ce{Cl^-}\}$ & 1--9 \\
        Alkane deprotonation &
        $\ce{C_n}\{\ce{COOH}\}\to
        \ce{C_n}\{\ce{COO^-}\}+\{\ce{H^+}\}$ & 1--9 \\
        Alkane hydrolysis &
        $\ce{C_n}\{\ce{COOH}\}+\{\ce{NH_3}\}\to
        \ce{C_n}\{\ce{CONH_2}\}+\{\ce{H_2O}\}$ & 1--9 \\
        \midrule
        Conjugated alkene substitution &
        $\ce{C_n}\{\ce{CH_2Cl}\}+\{\ce{OH^-}\}\to
        \ce{C_n}\{\ce{CH_2OH}\}+\{\ce{Cl^-}\}$ & 1--9 \\
        Conjugated alkene deprotonation &
        $\ce{C_n}\{\ce{COOH}\}\to
        \ce{C_n}\{\ce{COO^-}\}+\{\ce{H^+}\}$ & 3,5,7,9,11 \\
        Conjugated alkene hydrolysis &
        $\ce{C_n}\{\ce{COOH}\}+\{\ce{NH_3}\}\to
        \ce{C_n}\{\ce{CONH_2}\}+\{\ce{H_2O}\}$ & 3,5,7,9,11 \\
        \midrule
        Aromatic substitution &
        $\ce{Ar_n}\{\ce{Cl}\}+\{\ce{OH^-}\}\to
        \ce{Ar_n}\{\ce{OH}\}+\{\ce{Cl^-}\}$ & 6,10,14,16 \\
        Aromatic deprotonation &
        $\ce{Ar_n}\{\ce{COOH}\}\to
        \ce{Ar_n}\{\ce{COO^-}\}+\{\ce{H^+}\}$ & 6,10,14,16 \\
        Aromatic hydrolysis &
        $\ce{Ar_n}\{\ce{COOH}\}+\{\ce{NH_3}\}\to
        \ce{Ar_n}\{\ce{CONH_2}\}+\{\ce{H_2O}\}$ & 6,10,14,16 \\
        \midrule
        Diels--Alder (MVK + MeO-butadiene) &
        $\ce{CC(=O)}\{\ce{C=C}\}+\ce{CO}\{\ce{C(=C)C=C}\}
        \to\text{adduct}$ & 9 \\
        \midrule
        Thiol-Michael addition (cysteine + acrylamide) &
        $\ce{COOHC(N)}\{\ce{CSH}\}+\{\ce{C=C}\}\ce{CON}
        \to\text{adduct}$ & 6 \\
        \bottomrule
    \end{tabular*}
    
    \caption{Overview of the reaction dataset. Fragment atoms are in curly braces. The alkane and conjugated alkene series have linear carbon backbones; the aromatic series comprises benzene, naphthalene, anthracene, and pyrene derivatives. MVK denotes methyl vinyl ketone and MeO-butadiene denotes 2-methoxy-1,3-butadiene. Explicit reactant and product SMILES strings are provided in the Supporting Information (SI).}
    \label{tab:reactions}
\end{table*}

\subsection{Reaction dataset}
\label{subsec:dataset}

We evaluate CCSD(T)-in-HF calculations using the various active-space constructions discussed above for the prediction of 60 organic reaction energies, summarized in Table~\ref{tab:reactions}, including common reactions studied in previous embedding works~\cite{goodpaster_accurate_2014,claudino_simple_2019}. 
This set includes nine reaction families (e.g., alkane substitution, alkane deprotonation, etc.), each containing 4--9 individual reactions with a varying number of carbon atoms in the non-reactive part of the molecules, as well as one Diels-Alder reaction and one thiol-Michael addition. 
The studied reactions include cases where embedding is expected to be highly accurate (e.g., alkane substitution) and cases where embedding methods may be less effective [e.g., delocalized $\pi$ systems, systems with multiple reaction centers (Diels-Alder), and systems with strong non-local interactions (thiol-Michael)]. 
All systems are closed-shell.
    
We define the fragment to include only those atoms belonging to the reactive functional group.
These minimal fragments are explicitly highlighted in curly braces in Table~\ref{tab:reactions}.
The accuracy of quantum embedding can be improved by enlarging the size of the fragment, but here we are interestedin the performance for a fixed and conservative fragment size.
A primary motivation for studying reactions in molecules of varying chain lengths is to assess the size dependence of the embedding error.
An expanded version of Table~\ref{tab:reactions}, including explicit SMILES strings for all species in every reaction, is provided in the Supporting Information (SI).

\subsection{Threshold selection}

The active-space constructions described in Section~\ref{sec:theory} use fixed thresholds
to select the orbitals, which introduces several challenges to their black-box use and comparison.
First, as we will show below, fixed thresholds typically generate unbalanced active spaces across geometries; an especially problematic example of this occurs when the number of occupied orbitals is different, because the correlation energy is roughly proportional to the number of occupied orbitals.
Second, the different constructions do not necessarily generate active spaces of the same size as one another; because the active space size determines the cost of the CCSD(T) calculation, this complicates a fair comparison of accuracy versus cost.

Despite these two challenges, we have tried to select ``optimal'' thresholds for each method, which yield good performance and generate active spaces of similar sizes.
Because of the sharp drop-off in fragment populations, thresholds that vary by orders of magnitude typically yield active spaces that differ only by a handful of orbitals.
For Boys, PM, IBO, PAO, and DMET-RE orbitals, optimal thresholds were selected by testing on the hexane, hexene, and benzene substitution reactions.
We scanned all thresholds over several orders of magnitude and selected values that minimized the reaction energy error per frozen orbital; the scans are reported in the SI.
The selected thresholds in Table~\ref{tab:thresholds} were then used throughout our benchmark calculations.

\begin{table}[b]
    \centering
    \caption{Fixed thresholds for occupied and virtual active-space selection}
    \label{tab:thresholds}
    \renewcommand{\arraystretch}{1.2}
    \begin{tabular*}{\columnwidth}{@{\extracolsep{\fill}}lcc@{}}
        \toprule
        Method & Occupied threshold & Virtual threshold \\
        \midrule
        Boys & $0.01$ & $0.01$ \\
        PM / IBOs & $0.01$ & $0.1$ \\ 
        PAOs  & N/A & 0.005 \\ 
        DMET-RE & $0.25$ & $10^{-6}$ \\ 
        FragNOs & $10^{-4}$ & $10^{-5}$ \\
        \bottomrule
    \end{tabular*}
\end{table}

For FragNOs, we consider two different threshold schemes. 
In the first, we constrain the size of the occupied and virtual spaces to match those of DMET-RE, which allows us to directly quantify the per-orbital efficiency of FragNOs compared to those of DMET-RE. 
In the second scheme, we apply fixed thresholds (given in Tab.~\ref{tab:thresholds}), which were chosen by scanning over thresholds and identifying a plateau in the average dataset error.
The identified thresholds are similar to those typically used in LNO-CC~\cite{Rolik2011,Rolik2013,Nagy2024}, but slightly less conservative.

\subsection{Computational details}
\label{subsec:computational}

Initial molecular conformers were generated via the RDKit package and optimized in vacuum using the B3LYP-D3BJ\cite{becke_new_1993,grimme_effect_2011} functional and the def2-TZVP basis set with ORCA~\cite{neese_orca_2012,neese_software_2025}. 
For the hexane substitution and Diels-Alder reactions, minimum energy reaction paths were also generated in ORCA using the climbing-image nudged elastic band\cite{mills_reversible_1995,henkelman_improved_2000,asgeirsson_nudged_2021} (NEB) method at a B3LYP-D3BJ/def2-SVP level of theory. 

All subsequent calculations on these optimized geometries were performed using a development version of PySCF 2.6\cite{sun_pyscf_2018,sun_recent_2020}, which uses Libcint for Gaussian integral evaluation\cite{sun_libcint_2015}. 
PySCF calculations used the cc-pVTZ basis and the density-fitting approximation with the cc-pVTZ-JKFIT and cc-pVTZ-RI auxiliary basis sets.

All calculations were performed on a single node equipped with 32 CPU cores (Intel Xeon Gold 6226 2.9 Ghz) and 6 GB of memory per core.

\section{Results and Discussion}
\label{sec:results}

The accuracy of single-fragment CC-in-HF depends on both the orbital construction and the size of the retained space.
We first assess the occupied and virtual selection protocols specified in Section~\ref{sec:methods}, then use comparisons at matched dimensions to examine whether differences in accuracy persist at the same orbital count.
Finally, we assess correlation-based space enlargement and composite MP2 corrections, including their effects on reaction barriers.
Unless otherwise stated, errors are measured relative to full-system CCSD(T), and the family average is the equally weighted mean of the MAEs for the 11 reaction families.

\subsection{Occupied active spaces}
\label{subsec:occupied}
    \begin{table}[b]
        \caption{Mean absolute reaction-energy errors (kcal/mol) relative to full-system CCSD(T), using truncated occupied spaces and the full virtual space. The family average gives equal weight to the 11 reaction families. The ``unbalanced spaces'' row reports the family average results when the fixed occupation thresholds are used directly for reactant and product, without enforcing an equal number of occupied orbitals.
        }
        \label{tab:occ_error}
           \begin{tabularx}{\columnwidth}{@{}l*{4}{>{\centering\arraybackslash}X}@{}}
\toprule
& Boys & PM & IBO & DMET-RE \\
\midrule
Alkane substitution & 0.82 & 0.82 & 0.83 & 0.96 \\
Alkane deprotonation & 1.30 & 1.08 & 0.81 & 0.21 \\
Alkane hydrolysis & 0.72 & 2.57 & 2.78 & 0.37 \\
Alkene substitution & 2.94 & 0.86 & 0.43 & 0.80 \\
Alkene deprotonation & 0.98 & 1.75 & 3.81 & 2.92 \\
Alkene hydrolysis & 0.31 & 0.54 & 0.44 & 0.10 \\
Aromatic substitution & 1.75 & 3.15 & 3.53 & 0.71 \\
Aromatic deprotonation & 6.69 & 4.38 & 4.11 & 2.12 \\
Aromatic hydrolysis & 0.65 & 1.24 & 1.16 & 0.88 \\
Diels-Alder & 1.49 & 3.16 & 0.42 & 0.56 \\
Thiol-Michael & 6.65 & 6.69 & 6.65 & 7.42 \\
\midrule
Family average & {2.21} & {2.39} & {2.27} & {1.55} \\
Unbalanced spaces & 23.42 & 13.71 & 12.17 & 1.55 \\
\bottomrule
        \end{tabularx}
    \end{table}

We first compare Boys, PM, IBO, and DMET-RE occupied fragment spaces while retaining all virtual orbitals.
For reaction energies, the occupied selection must give a balanced treatment of reactants and products.
A difference in the number of correlated occupied orbitals can introduce a large imbalance in the recovered correlation energy, so consistency of these counts is an important starting condition for comparing the constructions~\cite{welborn_even-handed_2018,bensberg_direct_2019,bensberg_orbital_2020}, although we note that enforcing an equal number of orbitals does not ensure that the selected spaces describe the same correlation contributions.
To ensure an equal number of occupied orbitals, we apply our selected thresholds at the reactant geometry and then enforce the same number of occupied orbitals in the product geometry.

Table~\ref{tab:occ_error} reports the family-specific mean absolute errors (MAEs) for the occupied-space constructions tested.
With equal numbers of active occupied orbitals in reactants and products, all constructions give family-averaged MAEs below 2.5~kcal/mol, providing good accuracy at low computational cost.
As discussed in Section~\ref{ssec:comparing_fragments}, the semicanonicalized fragment orbitals obtained by all methods are quite similar, which explains their comparable performance (as long as the number of occupied orbitals is equal between reactants and products).

Applying fixed thresholds independently to reactants and products instead gives average errors of 10--50~kcal/mol for Boys orbitals, PM orbitals, and IBOs.
In contrast, we find that DMET-RE produces equal occupied-space dimensions for any threshold above 0.1, simplifying its use as a black-box method.
The family-average errors of this protocol are reported in the final row of Table~\ref{tab:occ_error}.

Overall, we find that DMET-RE gives the best performance, with most family MAEs below 1~kcal/mol, but alkene deprotonation, aromatic deprotonation, and cysteine--acrylamide adduct formation give errors of 2.92, 2.12, and 7.42~kcal/mol, respectively.
Thus, even with the full virtual space, the minimal-fragment occupied-orbital selection leaves important contributions to some reaction energies outside the correlated calculation.
These cases are understandably challenging due to their more delocalized electronic structure, which motivates enlarging the fragment or approximating the omitted correlation, the latter of which we explore further below.

\subsection{Virtual active spaces}
\label{subsec:virtual}

    \begin{table*}[htb]
        \caption{Mean absolute reaction-energy errors (kcal/mol) relative
to full-system CCSD(T), using truncated occupied spaces according to DMET-RE and truncated virtual spaces according to the methods indicated. For comparison, results from selected density functionals are also given.  
        }
        \label{tab:vir_error}
        \begin{tabularx}{\textwidth}{@{}l*{8}{>{\centering\arraybackslash}X}@{}}
        \toprule
                  & Boys & PM & PAO & FragPAO & DMET-RE (threshold) & DMET-RE (inflection) & B3LYP & $\omega$B97X-V \\
        \midrule
        Alkane substitution & 1.11 & 2.05 & 1.04 & 1.10 & 1.10 & 1.31 & 2.15 & 2.18 \\
        Alkane deprotonation & 3.91 & 5.26 & 2.98 & 1.03 & 1.02 & 5.30 & 2.94 & 1.33 \\
        Alkane hydrolysis & 0.75 & 1.12 & 1.19 & 0.55 & 0.55 & 0.86 & 0.07 & 0.07 \\
        Alkene substitution & 0.71 & 1.16 & 1.56 & 0.99 & 0.99 & 1.49 & 2.46 & 2.72 \\
        Alkene deprotonation & 1.17 & 3.72 & 1.23 & 1.73 & 1.73 & 3.73 & 3.09 & 1.66 \\
        Alkene hydrolysis & 2.40 & 0.23 & 0.78 & 0.09 & 0.09 & 0.08 & 0.32 & 0.13 \\
        Aromatic substitution & 0.70 & 1.45 & 0.53 & 0.24 & 0.24 & 0.31 & 3.27 & 3.07 \\
        Aromatic deprotonation & 0.77 & 4.93 & 0.71 & 0.67 & 0.67 & 5.13 & 1.93 & 1.10 \\
        Aromatic hydrolysis & 1.37 & 2.43 & 0.49 & 0.82 & 0.82 & 1.75 & 1.05 & 0.46 \\
        Diels-Alder & 0.49 & 0.53 & 0.59 & 0.21 & 0.19 & 1.26 & 12.96 & 6.39 \\
        Thiol-Michael & 7.34 & 9.26 & 9.26 & 6.99 & 7.72 & 6.47 & 9.38 & 3.59 \\
        \midrule
        Family average & {1.88} & {2.92} & {1.75} & {1.38} & {1.38} & {2.52} & {3.60} & {2.06} \\
        \bottomrule
        \end{tabularx}
    \end{table*}

\begin{figure}[b]
        \centering
        \includegraphics[width=\columnwidth]{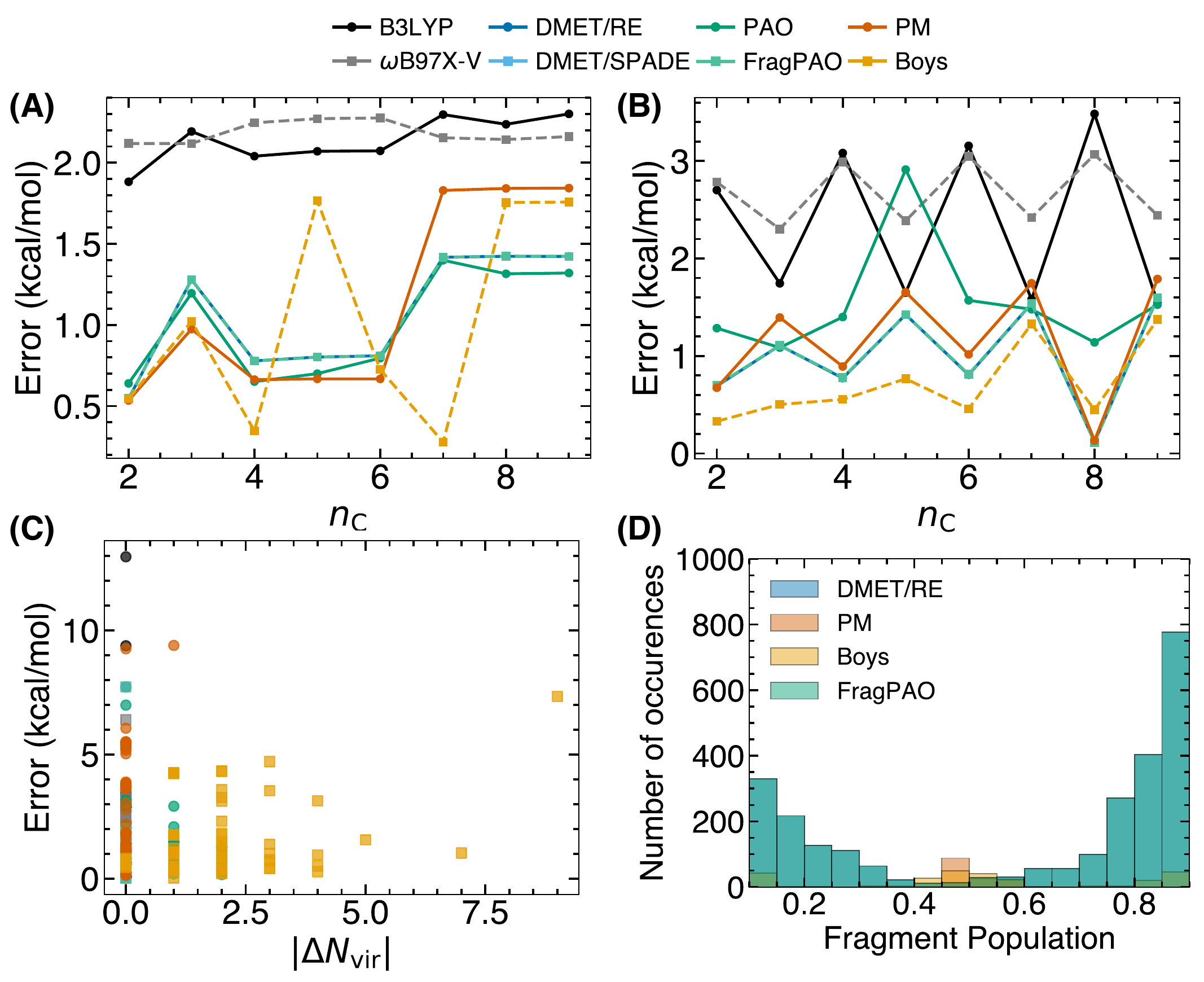}
        \caption{CC-in-HF error versus canonical CCSD(T) for virtual active-spaces. (A) Alkane substitution reaction errors. (B) Conjugated alkene substitution reaction errors. (C) Errors versus differences in the number of virtual orbitals in the active-space between reactants and products, $\Delta N_\text{vir}$. (D) Histograms for the number of orbitals with split populations (from 0.1--0.9) between fragment and environment.}
        \label{fig:vir_space}
    \end{figure}

We next fix the occupied space to that produced by DMET-RE and compare the virtual selection protocols, with results shown in Table~\ref{tab:vir_error};
in addition to DMET-RE with active orbital truncation by thresholds, we also present results obtained by truncating according to the largest inflection point in fragment populations (as suggested by the SPADE procedure~\cite{claudino_automatic_2019,claudino_simple_2019}).
At the thresholds specified in Table~\ref{tab:thresholds}, DMET-RE and FragPAOs give the smallest family-averaged MAEs, both approximately 1.38 kcal/mol, followed by PAOs, Boys, DMET-SPADE, and PM with MAEs of 1.75, 1.88, 2.52, and 2.92 kcal/mol, respectively (Table~\ref{tab:vir_error}).
Remarkably, for all of the virtual space methods, the errors presented within each family remain roughly constant across environments of different sizes, even when the environment contains a delocalized $\pi$-system, as can be seen in Figure~\ref{fig:vir_space}A and Figure~\ref{fig:vir_space}B.
And, unlike the case of occupied orbitals, differences in the number of correlated virtual orbitals have little affect on the error (Figure~\ref{fig:vir_space}C). 
The SPADE selection protocol typically identifies a smaller number of virtual orbitals than the RE protocol with fixed thresholds, which explains the former's worse accuracy.
For comparison, density functional theory results are also reported for the B3LYP and $\omega$B97X-V functionals, which yield family-averaged MAEs of 3.60~kcal/mol and 2.06~kcal/mol, respectively.

In general, DMET-RE and FragPAOs are the best performing active-space virtual orbitals, with nearly identical average errors of 1.38~kcal/mol. The near equivalence of these numerical results can be shown to be a consequence of the close mathematical relationship between the two methods.
Interestingly, the DMET-RE error is slightly lower when both the occupied and virtual spaces are truncated than when only the occupied space is, suggesting a fortuitous cancellation of errors. 
As in the previous section, most methods---including DMET-RE and FragPAOs---struggle with the deprotonation reactions and the thiol-Michael addition. 

Unlike for occupied orbitals, where the separation between fragment and environment orbitals is relatively clear, we find that most methods lead to a large number of orbitals with intermediate fragment populations, e.g., between 0.1 and 0.9, as shown in Figure~\ref{fig:vir_space}D. 
Including these delocalized orbitals is crucial for obtaining accurate results, which is why the virtual orbital thresholds are so much tighter than the occupied orbital thresholds. 
These finding are consistent with conclusions drawn in Refs.~\citenum{barcza_benchmarking_2023} and \citenum{szirmai_projected_2024}, which illustrated that, when embedding on a single monomer in a dimer complex, the virtual active-space must extend across entire dimer to ensure accurate results.

\begin{figure}[b]
        \centering
        \includegraphics[width=\columnwidth]{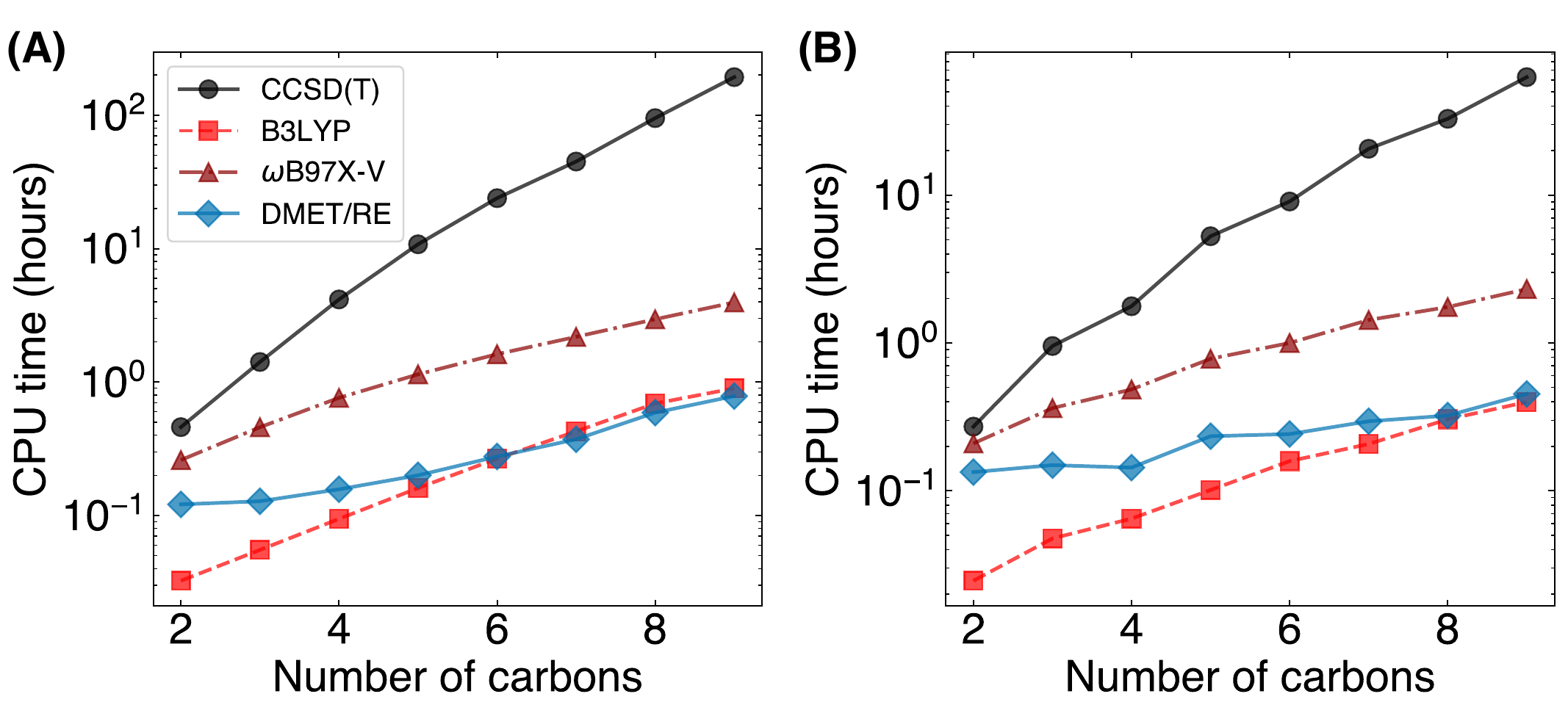}
        \caption{Computational cost in terms of CPU hours for full-system CCSD(T), B3LYP, $\omega$B97X-V, and CCSD(T)-in-HF using DMET-RE. (A) Alkane substitution reaction. (B) Alkene substitution reaction.}
        \label{fig:cost}
    \end{figure}

With a few specific exceptions, our results have shown that 1.5~kcal/mol accuracy or better is possible with minimal fragment embedding and a simple CCSD(T)-in-HF energy expression.
Importantly, these calculations are extremely cheap and scale favorably with total system size; they are comparable in cost to hybrid functionals, cheaper than range-separated hybrid functionals (as implemented in PySCF), and significantly cheaper than full CCSD(T), as shown in Fig.~\ref{fig:cost}.
Improved results can be obtained by increasing the size of the fragment, although this procedure is not black-box, and we find that difficult test cases require fragments that span nearly the entire molecule in order to reach sub-kcal/mol accuracy (see SI).
Another option is to improve the description of electron correlation in the environment, such as by using second-order perturbation theory, which we test in the next section.

\subsection{MP2 fragment natural orbitals and energy corrections}
\label{subsec:mp2}

The fragment-based active-space methods studied in the previous section only have a few steps that scale with the total system size $N$: the initial HF calculation, which scales formally as $N^4$ but smaller in practice; the evaluation and rotation of the three-center density-fitting integrals, both of which scale formally as $N^3$, which can be reduced with domains or local density fitting.
However, there are many situations where a second-order M\o ller-Plesset perturbation theory (MP2) calculation may be feasible, while a higher-level calculation like CCSD(T) may not be. 
In this section, we explore how improvements from MP2 theory---either in the construction of the active-space orbitals via FragNOs (as described in Sec.~\ref{subsec:fragnos}) or in the energy correction of Eq.~\ref{eq:delmp2}---affect the accuracy of the embedding methods.

\begin{table}[b]
    \caption{Mean absolute error (kcal/mol) by reaction family and averaged over reaction families for methods without and with MP2 energy corrections ($\Delta E^{(2)}$). FragNO results use either a threshold to truncate the virtual space or a number of virtual orbitals that is matched to that of DMET-RE.}
    \label{tab:mp2_error}
    \begin{tabularx}{\columnwidth}{@{}l*{4}{>{\centering\arraybackslash}X}@{}}
    \toprule
    & \multicolumn{2}{c}{without $\Delta E^{(2)}$}
& \multicolumn{2}{c}{with $\Delta E^{(2)}$} \\
    & FragNO (thresh.) & FragNO (matched) & DMET-RE & FragNO (thresh.)\\
    \midrule
    Alkane substitution & 0.38 & 1.27 & 0.38 & 0.09 \\
    Alkane deprotonation & 0.54 & 1.38 & 0.10 & 0.13 \\
    Alkane hydrolysis & 0.04 & 0.50 & 0.03 & 0.03 \\
    Alkene substitution & 0.49 & 0.86 & 0.31 & 0.15 \\
    Alkene deprotonation & 1.42 & 1.28 & 0.46 & 0.63 \\
    Alkene hydrolysis & 0.06 & 0.56 & 0.11& 0.08 \\
    Aromatic substitution & 1.52 & 1.18 & 1.24 & 0.50 \\
    Aromatic deprotonation & 1.20 & 0.39 & 0.39 & 0.69 \\
    Aromatic hydrolysis & 0.23 & 1.31 & 0.48 & 0.47 \\
    Diels-Alder & 0.62 & 0.38 & 0.39 & 0.17 \\
    Thiol-Michael & 4.50 & 8.06 & 0.23 & 0.25 \\
    \midrule
    Family average & 1.00 & 1.56 & 0.37 & 0.29 \\
    \bottomrule
    \end{tabularx}
\end{table}

Our results are summarized in Table~\ref{tab:mp2_error}.
The second column of Table~\ref{tab:mp2_error} presents the MAEs for each of the reaction types for FragNOs, using DMET-RE occupied orbitals to define the fragment MP2 matrix.
We find that FragNO virtual orbitals outperform those from DMET-RE for nearly every reaction type, however much of this improved performance is a consequence of the larger virtual active-space selected by the FragNO scheme with our recommended threshold.
If one restricts FragNOs to have the same number of occupied and virtual active orbitals as DMET-RE, the average error is worse than DMET-RE. 
Therefore, the per-orbital efficiency of FragNOs is worse than DMET-RE on average.
This result suggests that locality is more important than correlation, at least when calculating energy differences of chemical reactions with local modifications.

A much more significant improvement can be obtained by simply using an MP2 composite correction to the energy (Eq.~\ref{eq:delmp2}), as shown in the final three columns of Table~\ref{tab:mp2_error}. 
With these corrections, the errors of all methods tested are well within chemical accuracy, with MAEs of 0.37 kcal/mol for both DMET-RE virtuals and FragPAO virtuals, and an MAE of 0.29 kcal/mol for FragNO virtuals. 

\subsection{Reaction paths}

As a final demonstration, we test several active-space construction methods on the prediction of energies along an entire reaction path; the reaction path geometries were found via climbing image NEB with B3LYP+D3BJ functional (see Sec.~\ref{subsec:computational}.
We performed these calculations on the hexane substitution reaction, the Diels-Alder reaction, and thiol-Michael addition (for this last reaction, we deprotonated the cysteine thiol group in the reactant geometry, as this reaction is catalyzed under basic conditions). 

    \begin{figure*}[htb]
        \centering
        \includegraphics[width=0.9\textwidth]{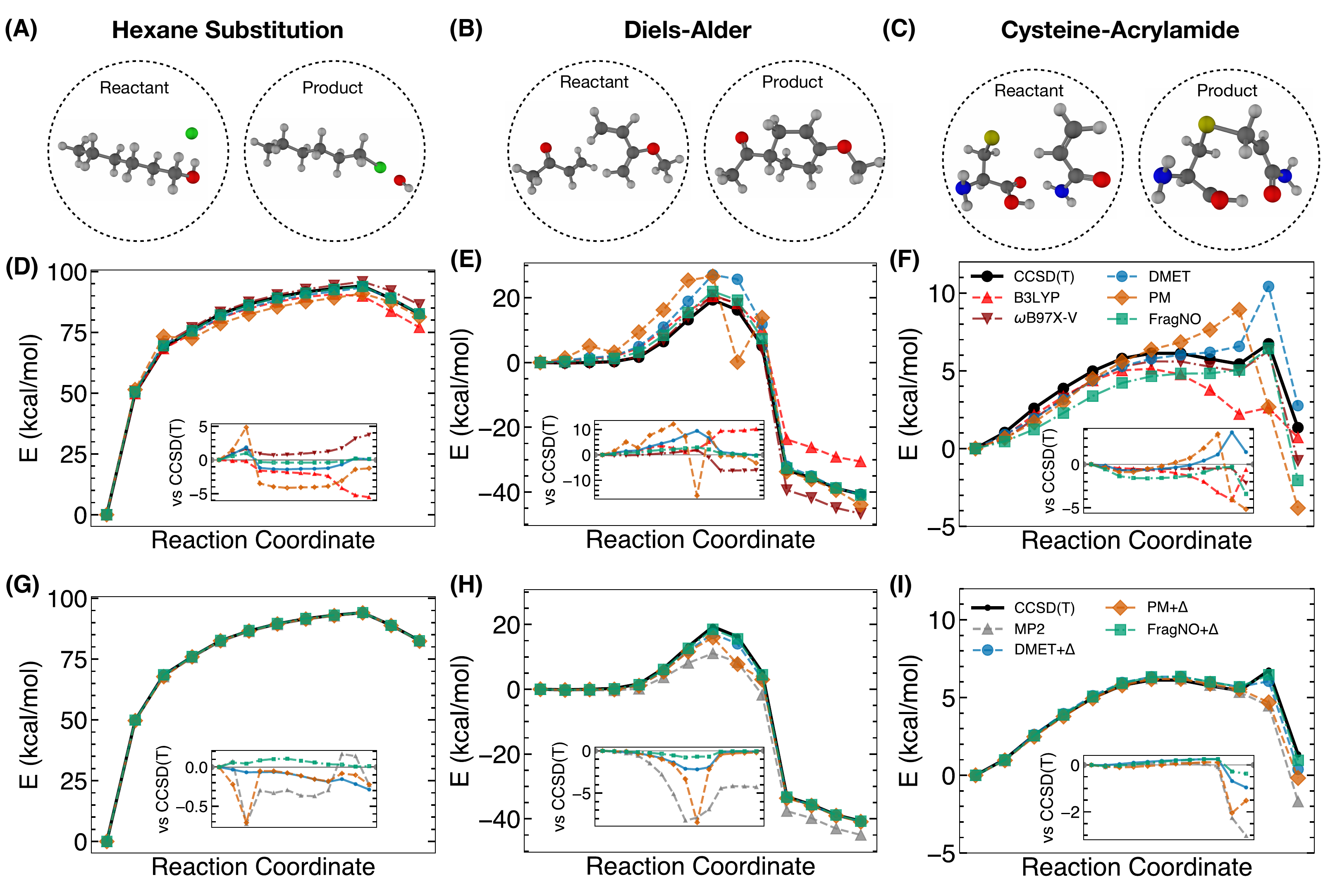}
        \caption{Performance of CCSD(T)-in-HF embedding compared to full-systel CCSD(T) for reaction path energies for the hexane substitution, the MVK+MeO-butadiene Diels-Alder reaction, and the cysteine+ acrylamide thiol-Michael addition.
        Panels in the top row [(A), (B), and (C)] compare methods without MP2 energy corrections, while panels in the bottom row [(D), (E), and (F)] compare methods with MP2 energy corrections.}
        \label{fig:neb}
    \end{figure*}
    
Results are shown in Figure~\ref{fig:neb}, which compares reference CCSD(T) reaction profiles to those of CCSD(T)-in-HF using several active-space construction methods without the MP2 energy correction (A, B, C) and with the MP2 energy correction (D, E, F).
Importantly, the results obtained using DMET-RE and FragNO active spaces do not display any significant discontinuities, reflecting the balanced orbital selection provided by these methods.
In contrast, the use of fragment PM orbitals does exhibit several large discontinuities. 

For the hexane substitution reaction, CCSD(T)-in-HF embedding with DMET-RE is sufficient to capture the entire MES at high accuracy, significantly outperforming DFT without added cost, even without MP2 energy corrections. 
The localized electronic structure of alkane molecules and the SN$_2$ reaction are well-suited to a robust single-fragment embedding approach. 
    
In contrast, for the Diels-Alder reaction, CCSD(T)-in-HF exhibits worse performance. 
While one-shot DMET-RE is sufficient for the total reaction energy, it overestimates the barrier height by almost 10 kcal/mol (Figure~\ref{fig:neb}B). 
These errors arise because of steric shifts in the molecular sidechains, which are not included in the fragment. 
The use of FragNOs reduces the error in the barrier height to only 3~kcal/mol, but this is significantly worse than that predicted by B3LYP or $\omega$B97X-V (although the DFT reaction energy errors are significantly larger). 
The addition of MP2 composite corrections (Figure~\ref{fig:neb}E) restores the accuracy of the barrier height to about 2~kcal/mol for DMET-RE and about 1~kcal/mol for FragNOs. 

Finally, for the thiol-Michael addition (the hardest example in our dataset), results obtained with DMET-RE and FragNO active spaces exhibit errors of almost 4~kcal/mol near the product geometry, but the MP2 energy correction restores the accurate to better than 1~kcal/mol. Once again, the addition of MP2 composite corrections restores the accuracy of the barrier height to within 1 kcal/mol
Taken together, these results highlight the value of treating the environment using higher levels of theory than HF. 
    
\section{Conclusion}
We have discussed and systematically tested the performance of various single-fragment active-space selection methods, many of which are commonly used in quantum embedding theories. 
For the active occupied orbitals, we have confirmed that maintaining a balanced active-space size, (i.e., one that keeps the same number of active orbitals between different geometries) is essential to minimizing error; although this can be enforced by hand, we found that DMET-RE typically produces such balanced active spaces for reasonable choices of the population threshold.

For the active virtual orbitals, we found that both DMET-RE and FragPAOs performed the best.
These methods had the greatest number of orbitals with populations delocalized over fragment and the environment, which we hypothesize are important for capturing important fragment-environment correlations. 
When using DMET-RE for both occupied and virtual fragment orbitals, or DMET-RE for occupied and FragPAOs for virtual orbitals, CCSD(T)-in-HF embedding outperforms common DFT functionals such as B3LYP and $\omega$B97X-V---providing improved accuracy at comparable cost.
However, the average embedding errors are still about 1.5~kcal/mol, and the most difficult cases having errors greater than 5~kcal/mol.
By incurring the increased cost of a global MP2 calculation, we showed that FragNOs provide competitive virtual space orbitals and that MP2 energy corrections significantly reduce the embedding errors.
Several of our best performing methods, along with our recommended fragment population thresholds, offer nearly black-box methods that provide nearly CCSD(T) accuracy at either HF or MP2 cost.
At lower cost than that of MP2, accurate density functionals could be used for composite energy corrections~\cite{manby_simple_2012,goodpaster_accurate_2014,Lee2019}, which will be explored in future work.

\acknowledgments

We thank Garnet Chan and Hong-Zhou Ye for helpful discussions.
This work was supported by the Columbia Center for Computational Electrochemistry.
The Flatiron Institute is a division of the Simons Foundation.

\end{document}